\documentclass[aps,pra,twocolumn,superscriptaddress]{revtex4-1}
\usepackage{amssymb}
\usepackage{graphicx}
\usepackage{amsmath}
\usepackage{mathrsfs}
\usepackage{bm}
\usepackage{dcolumn}
\usepackage[dvipdfm]{hyperref}
\begin{document}
\title{Sudden transition between classical and quantum decoherence in dissipative cavity QED and stationary quantum discord}
\author{Qi-Liang He}
\affiliation{Zhejiang Institute of Modern Physics and Physics Department, Zhejiang
University, Hangzhou 310027, People's Republic of China}
\author{Jing-Bo Xu}
\email{xujb@zju.edu.cn}
\affiliation{Zhejiang Institute of Modern Physics and Physics Department, Zhejiang
University, Hangzhou 310027, People's Republic of China}
\author{Dao-Xin Yao}
\affiliation{State Key Laboratory of Optoelectronic Materials and Technologies, School of Physics and Engineering, Sun Yat-sen University, Guangzhou 510275, People's Republic of China}
\affiliation{Department of Physics and Astronomy, University of Tennessee, Knoxville, TN 37996, USA}
\author{Ye-Qi Zhang}
\affiliation{Zhejiang Institute of Modern Physics and Physics Department, Zhejiang
University, Hangzhou 310027, People's Republic of China}
\date{\today}

\begin{abstract}
We investigate the phenomenon of sudden transition between classical and quantum decoherence in the study of quantum discord for a dissipative cavity QED system, which consists of two noninteracting two-level atoms, each trapped in a dissipative cavity. It is found that the quantum discord between two atoms, which are prepared initially in the X-type quantum states, is not destroyed by the dissipation of the cavities for a finite time interval and the stationary quantum discord can arise in the interaction of atoms with cavities as the time approach to infinite. The transition time is sensitive to the initial state parameter of the two atoms and the mean photon number of the coherent field. Interestingly, the quantum discord between the two atoms is completely unaffected by the dissipation of the cavities if we choose the suitable value of the ratio, which depends on the decay rate of two cavities and the atom-field coupling constant.
\end{abstract}
\maketitle
\section{Introduction}
Quantum correlations play an important role in quantum information and computation and have become a hot topic of intense research in recent years. Entanglement is a special kind of quantum correlation and has been recognized as an essential resource for many operations in quantum information processing \cite{1,2,3,4}. However, the entanglement is not the only type of quantum correlation and there exist quantum tasks that display the quantum advantage without entanglement \cite{5,6,7}. It has been demonstrated both theoretically \cite{8,9,10} and experimentally \cite{11} that other nonclassical correlation, namely, quantum discord \cite{12} can be responsible for the computational speedup for certain quantum tasks. Quantum discord, introduced in Ref.~\cite{12}, is defined as the difference between the quantum mutual information and the classical correlation and is nonzero even for separate mixed states. Therefore, the quantum discord may be regarded as a more general and fundamental resource for quantum information processing.

On the other hand, a real quantum systems will unavoidably be influenced by
surrounding environments. The interaction between the quantum system and its
environment leads to a rapid destruction of quantum coherence, which is
the main problem for the realization of quantum information processing.
Therefore, it is very important to understand the dynamics of quantum correlations of open
systems and find potential applications in quantum information theory. Many efforts have been devoted to the study of the dynamics of
quantum correlation under various decoherence channels \cite{13,14,15,16}. It has been shown that the discord is more robust than entanglement under the Markovian environments \cite{14} and vanishes only at some time points under non-Markovian environment \cite{16}.

Recently, the dynamics of quantum and classical correlations have been studied in the presence of nondissipative decoherence and the phenomenon of the sudden transition from classical to quantum decoherence in a finite time interval has been reported \cite{17}. Some of these phenomena have been observed in the recent experiment \cite{18}. In this paper, we investigate the phenomenon of sudden transition between classical and quantum decoherence in the study of quantum discord for the dissipative cavity QED system, which consists of two noninteracting two-level atoms, each trapped in a dissipative cavity. Firstly, we assume that the two atoms are prepared initially in the X-type quantum states and show that the quantum discord of two atoms is not destroyed by the dissipation of the cavities for a finite time interval and revivals to a stable value after damping oscillation, which means that there is a stationary quantum discord between the two atoms as the time approach to infinite. This implies that the initial quantum discord of two atoms can be partially preserved even when they are put into the two spatially separated dissipative cavities, respectively. Furthermore, we can see clearly that the transition time depends on the parameter of the initial states of the two atoms and the mean photon number of the coherent field. Particularly, it is interesting to point out that the quantum discord of two atoms is completely unaffected by the dissipation of the cavities if we choose the suitable ratio, which depends on the decay rate of two cavities and the atoms-field coupling constant. Then, we find that the sudden transition phenomenon does not appear in this system if the two atoms are initially in the Werner state. Instead, the stationary quantum discord of the two atoms still exists in the long-time regime. It is worth noting that the amount of the stationary quantum discord between two atoms can be enhanced by increasing the value of the ratio.
\section{The dynamics of two atoms in dissipative cavities}

\begin{figure}
\centering
\includegraphics[width=6cm]{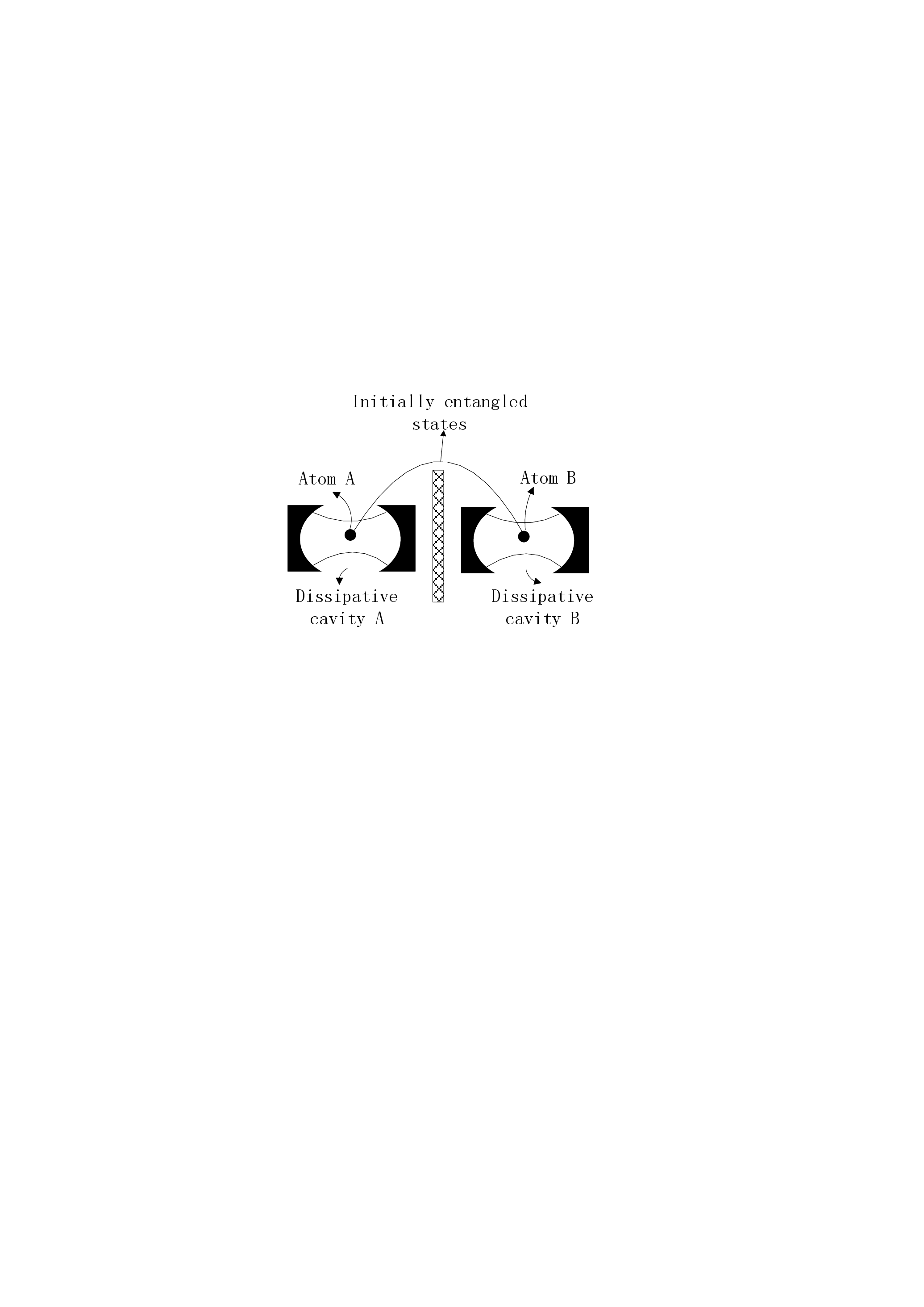}
\caption{\label{fig:fig1} This is the schematic diagram of the system which is investigated
in the present paper. There is no interaction between two atoms and no
communication between two cavities.}
\end{figure}

We consider the system consisting of two noninteracting two-level atoms, each trapped inside a dissipative cavity (see Fig.~\ref{fig:fig1}). The
Hamiltonian describing the interaction between atoms and cavities can be
written as ($\hbar =1$)
\begin{equation}{\label{Eq.1}}
H=\frac{\omega _{0}}{2}(\sigma _{A}^{z}+\sigma
_{B}^{z})+\omega (a_{A}^{\dagger }a_{A}+b_{A}^{\dagger
}b_{A})+g\sum_{i=A,B}(a_{i}^{\dagger }\sigma _{i}^{-}+a_{i}\sigma
_{i}^{+}),
\end{equation}
where $g$ is the atom-field coupling constant, $a^{\dagger }$ and $a$ are the creation and annihilation operator of the single-mode cavity field, and $\sigma^z$ is the atomic inversion operator, $\sigma _{i}^{+}=|e\rangle_i\langle g|$ ($\sigma _{i}^{-}=|g\rangle_i\langle e|$) is the atomic spin flip operators. The symbols $|e\rangle$ and $|g\rangle$ refer to the excited and ground states of the two-level atom. $\omega_0$ and $\omega$ are the atomic transition frequency and cavity frequency, respectively. It is clear that there is no interaction between subsystem A and subsystem B, which means that the evolution of each subsystem is independent. In the dispersive approximation, the Hamiltonian of Eq.~(\ref{Eq.1}) can be rewritten as \cite{19}
\begin{eqnarray}{\label{Eq.2}}
H_{\textrm{eff}}&=&\frac{\omega _{0}}{2}(\sigma _{A}^{z}+\sigma
_{B}^{z})+\omega (a_{A}^{\dagger }a_{A}+b_{A}^{\dagger
}b_{A})+\Omega[(a_{A}^{\dagger }a_{A}+1)\nonumber\\
&&|e\rangle_A\langle e|-a_{A}^{\dagger }a_{A}|g\rangle_A\langle g|]+\Omega[(a_{B}^{\dagger }a_{B}+1)\nonumber\\
&&|e\rangle_B\langle e|-a_{B}^{\dagger }a_{B}|g\rangle_B\langle g|],
\end{eqnarray}
with $\Delta=\omega_0-\omega$ and $\Omega=g^2/\Delta$.

Next, we investigate the dynamical evolution of two noninteracting two-level atoms interacting with two dissipative cavities by making use of the master equation. In the dispersive approximation, the master equation can be read as
\begin{eqnarray}{\label{Eq.3}}
\frac{d\rho(t)}{dt}&=&-i[H_{\textrm{eff}},\rho(t)]+D\rho(t)\nonumber\\
& &-i[H_{\textrm{eff}},\rho(t)]+D_A\rho(t)+D_B\rho(t),
\end{eqnarray}
where the superoperators $D_A$ and $D_B$ represent the losses in the cavity A and B. At zero temperature, the superoperators $D_A$ and $D_B$ can be written as
\begin{eqnarray}
D_A&=&k(2a_A\cdot a^\dagger_A-a^\dagger_A a_A\cdot-\cdot a^\dagger_A a_A),\nonumber\\
D_B&=&k(2a_B\cdot a^\dagger_B-a^\dagger_B a_B\cdot-\cdot a^\dagger_B a_B),
\end{eqnarray}
where $k$ is the cavity decay rate. For simplicity, we assume the decay rates of two cavities are equal. In the interaction picture, the master equation takes the form
\begin{equation}{\label{Eq.5}}
\frac{d\rho(t)}{dt}=-i[V,\rho(t)]+D_A\rho(t)+D_B\rho(t),
\end{equation}
with
\begin{equation}
V=\Omega\sum_{i=A,B}[(a_{i}^{\dagger }a_{i}+1)|e\rangle_i\langle e|-a_{i}^{\dagger }a_{i}|g\rangle_i\langle g|].
\end{equation}

This equation can be solved by using the superoperator method \cite{19,20,21}. We assume  that the two cavities are prepared initially in the coherent states $|\alpha_1\rangle$ and $|\alpha_2\rangle$, and the two atoms are prepared in a class of state with the maximally mixed marginals, which can be described by the three-parameter X-type density matrix in the basis $\{|ee\rangle,|eg\rangle,|ge\rangle,|gg\rangle\}$ as
\begin{eqnarray}
&&\rho_{\textrm{atom}}(0)=\frac{1}{4}(I+\sum_{i=1}^{3}c_{i}\sigma_{A}^{i}\otimes\sigma_{B}^{i})\nonumber\\
& &=\frac{1}{4}\left(\begin{array}{cccc}
1+c_{3} & 0 & 0 & c_{1}-c_{2} \\
0 & 1-c_{3} & c_{1}+c_{2} & 0 \\
0 & c_{1}+c_{2} & 1-c_{3} & 0 \\
c_{1}-c_{2} & 0 & 0 & 1+c_{3}
\end{array} \right),
\end{eqnarray}
where $c_i$ ($0\leq|c_i|\leq1)$ are the real numbers , $I$ is the identity operator of the total system, $\sigma_{A}^{i}$ and $\sigma_{B}^{i}$ are the Pauli operators of atom A and atom B, respectively. For simplicity, we set $\alpha_1=\alpha_2=\alpha$. Then, the density matrix of the whole system at $t=0$ is
\begin{eqnarray}
\rho(0)&=&\frac{1}{4}\left(\begin{array}{cccc}
1+c_{3} & 0 & 0 & c_{1}-c_{2} \\
0 & 1-c_{3} & c_{1}+c_{2} & 0 \\
0 & c_{1}+c_{2} & 1-c_{3} & 0 \\
c_{1}-c_{2} & 0 & 0 & 1+c_{3}
\end{array} \right) \nonumber\\
& &\otimes |\alpha\rangle_A\langle\alpha| \otimes |\alpha\rangle_B\langle\alpha|.
\end{eqnarray}

It is not difficult to find that the density matrix of the system at time $t$ is
\begin{eqnarray}{\label{Eq.9}}
&&\rho(t)=\frac{1+c_3}{4}|ee\rangle_{AB}\langle e e||\alpha_+(t)\alpha_+(t)\rangle_{AB}\langle\alpha_+(t)\alpha_+(t)|\nonumber\\
&&+\frac{1-c_3}{4}|eg\rangle_{AB}\langle eg||\alpha_+(t)\alpha_-(t)\rangle_{AB}\langle\alpha_+(t)\alpha_-(t)|+\nonumber\\
&&\frac{1-c_3}{4}|ge\rangle_{AB}\langle ge||\alpha_-(t)\alpha_+(t)\rangle_{AB}\langle\alpha_-(t)\alpha_+(t)|+\frac{1+c_3}{4}\nonumber\\
&&|gg\rangle_{AB}\langle g g||\alpha_-(t)\alpha_-(t)\rangle_{AB}\langle\alpha_-(t)\alpha_-(t)|+\{\frac{c_1-c_2}{4}f(t)^2\nonumber\\
&&|ee\rangle_{AB}\langle g g||\alpha_+(t)\alpha_+(t)\rangle_{AB}\langle\alpha_-(t)\alpha_-(t)|+\frac{c_1+c_2}{4}|f(t)|^2\nonumber\\
&&|eg\rangle_{AB}\langle ge| |\alpha_+(t)\alpha_-(t)\rangle_{AB}\langle\alpha_-(t)\alpha_+(t)|+H.C.\},
\end{eqnarray}
with
\begin{eqnarray}
f(t)&=&\exp\{-i\Omega t+|\alpha|^2(e^{-2kt}-1)\}\nonumber\\
& &\cdot\exp\{\frac{|\alpha|^2k}{k+i\Omega}(1-e^{-2(k+i\Omega)t})\},\nonumber\\
|\alpha_\pm(t)\rangle&=&|\alpha \exp{-(k\pm i\Omega)t}\rangle.
\end{eqnarray}
where $H.C.$ represents the Hermitian conjugate.
\section{The dynamics of quantum discord of two atoms}

In this section, we investigate the dynamics of quantum discord of two atoms which are trapped in the two spatially separated and dissipative cavities, respectively. For a two-qubit quantum system, the quantum discord qualifying a measure of quantum correlation, is defined as the difference between the quantum mutual information and the classical correlation,
\begin{equation}
\mathcal{Q} (\rho_{AB}) =\mathcal{I}(\rho_{AB}) -
\mathcal{C}(\rho_{AB}),
\end{equation}
where $\mathcal{I}(\rho_{AB})$ is the total correlation of two subsystem and can be expressed as
\begin{equation}
\mathcal{I}(\rho_{AB})=S(\rho_{A}) + S(\rho_{B}) -
S(\rho_{AB}),
\end{equation}

Here, $S(\rho) = -tr(\rho\log_2{\rho})$ is the von Neumann entropy, $\rho_A$ and $\rho_B$ are the reduced density matrices of $\rho_{AB}$. Besides, $\mathcal{C}(\rho_{AB})$ is the classical correlation between two subsystem A and B, which is defined as the maximum information one can obtain form A  by performing a perfect measurement on B. As discussed in Ref.~\cite{22}, the classical correlation is described as
\begin{equation}{\label{eq.13}}
\mathcal{C}(\rho_{AB})=\max_{\{B_{k}\}}\{S(\rho_{A}) -
S(\rho_{AB}|{\{B_{k}\}}),
\end{equation}
where $\{B_{k}\}$ is a complete set of projectors preformed locally on
subsystem B. $S(\rho_{AB}|\{B_{k}\})=\sum_k p_kS(\rho_{k})$
is the based-on-measurement quantum conditional entropy,
$(\rho_{AB}|\{B_{k}\} ):=\rho_k=1/p_k (I\otimes B_k) \rho_{AB}
(I\otimes B_k)$ is the conditional density operator and
$p_k=tr_{(AB)}[(I\otimes B_k) \rho_{AB} (I\otimes B_k)]$ is the
probability.

Next, we begin to study the quantum discord dynamic properties of two atoms. Tracing over the degrees of the freedom of cavity fields in the Eq.~(\ref{Eq.9}), we can obtain the reduced density matrix of atom A and atom B,
\begin{eqnarray}{\label{eq.14}}
&&\rho_{AB}(t)=\frac{1+c_3}{4}|ee\rangle\langle ee|+  \frac{1-c_3}{4}|eg\rangle\langle eg|+\frac{1-c_3}{4}\nonumber\\
& &|ge\rangle\langle ge|+\frac{1+c_3}{4}|gg\rangle\langle gg|+\{\frac{c_1-c_2}{4}(f(t)\chi(t))^2\nonumber\\
& &|ee\rangle\langle gg|+\frac{c_1+c_2}{4}|f(t)\chi(t)|^2|eg\rangle\langle ge|+H.C.\},
\end{eqnarray}
where $\chi(t)=\langle\alpha_-(t)|\alpha_+(t)\rangle$ and $H.C.$ is the Hermitian conjugate.

The eigenvalues of the reduced density matrix $\rho_{AB}(t)$ in Eq.~(\ref{eq.14}) can be derived as,
\begin{eqnarray}{\label{eq.15}}
\lambda_{1,2}&=&\frac{1}{4}[(1-c_3)\pm|f(t)\chi(t)|^2(c_1+c_2)],\nonumber\\
\lambda_{3,4}&=&\frac{1}{4}[(1+c_3)\pm|f(t)\chi(t)|^2(c_1-c_2)].
\end{eqnarray}

It is not difficult to find from Eq.~(\ref{eq.14}) that the reduced density matrix of each subsystem is maximally mixed, which means that $\rho_A(t)=\rho_B(t)=I/2$. Consequently, the von Neumann entropy $S(\rho_A(t))=S(\rho_B(t))=1$. Then the quantum mutual information between two subsystems is
\begin{equation}{\label{eq.16}}
I(\rho_{AB}(t))=2+\sum_{i=1}^4\lambda_i\log_2\lambda_i.
\end{equation}

In order to calculate the classical correlation $\mathcal{C}(\rho_{AB})$, we choose the complete set of projectors $\{B_k=|\theta_k\rangle\langle\theta_k|,$ $(k=1,2)\}$ to measure the subsystem B, where the two orthogonal projectors are defined by,
\begin{eqnarray}
|\theta_1\rangle&=&\cos\theta|e\rangle+e^{i\phi}\sin\theta|g\rangle,\nonumber\\
|\theta_2\rangle&=&e^{-i\phi}\sin\theta|e\rangle-\cos\theta|g\rangle,
\end{eqnarray}
with the parameters $\theta$ and $\phi$ vary from $0$ to $2\pi$, respectively. After the measurement $\{B_k,(k=1,2)\}$, the probability $p_1=p_2=1/2$ and the reduced matrices of subsystem A can be obtained as follows,
\begin{eqnarray}{\label{eq.18}}
&&\rho^1_A(t)=\frac{1}{2}[(1+c_3)\cos^2\theta+(1-c_3)\sin^2\theta]|e\rangle_A\langle e|\nonumber\\
&&+\frac{1}{2}[(1-c_3)\cos^2\theta+(1+c_3)\sin^2\theta]|g\rangle_A\langle g|+\nonumber\\
&&\frac{1}{2}\{[(c_1-c_2)e^{i\phi}f^2(t)\chi^2(t)\cos\theta\sin\theta+(c_1+c_2)\nonumber\\
&&e^{-i\phi}|f(t)\chi(t)|^2\cos\theta\sin\theta]|e\rangle_A\langle g|+H.C.\},\nonumber\\
&&\rho^2_A(t)=\frac{1}{2}[(1-c_3)\cos^2\theta+(1+c_3)\sin^2\theta]|e\rangle_A\langle e|\nonumber\\
&&+\frac{1}{2}[(1+c_3)\cos^2\theta+(1-c_3)\sin^2\theta]|g\rangle_A\langle g|+\nonumber\\
&&\frac{1}{2}\{[-(c_1-c_2)e^{i\phi}f^2(t)\chi^2(t)\cos\theta\sin\theta-(c_1+c_2)\nonumber\\
&&e^{-i\phi}|f(t)\chi(t)|^2\cos\theta\sin\theta]|e\rangle_A\langle g|+H.C.\},
\end{eqnarray}
where $H.C.$ denotes the Hermitian conjugate.
\begin{figure}
\centering
\includegraphics[width=7cm]{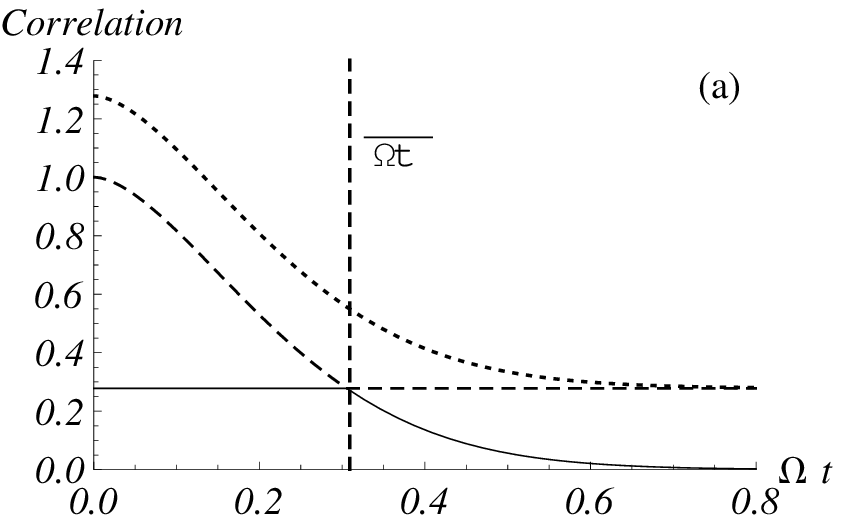}
\includegraphics[width=7cm]{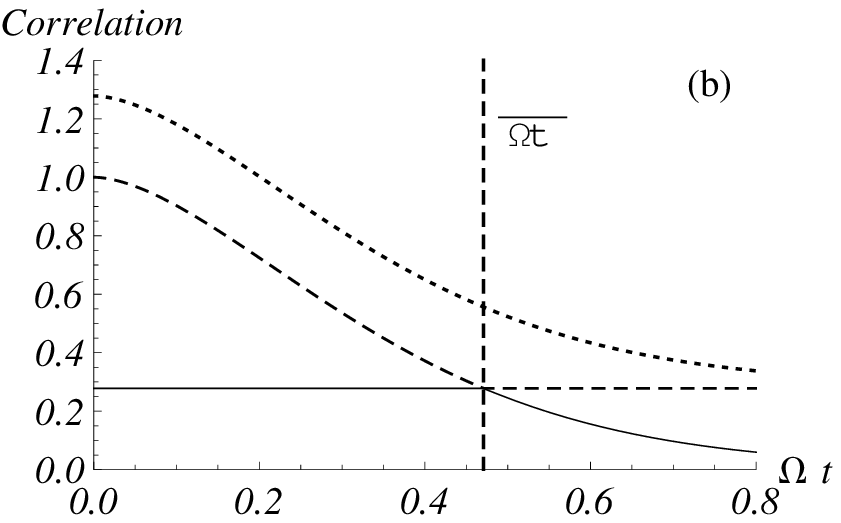}
\caption{\label{fig:fig2} Time evolution of quantum mutual information (dotted line), classical correlation (dashed line) and quantum discord (solid line) of two atoms as a function of the dimensionless scaled time $\Omega t$ with $k/\Omega=0.05$ and $c_3=0.6$. Vertical dashed line corresponds to the transition time $\overline{\Omega t}$. Panel (a): $\alpha=1.2$. Panel (b): $\alpha=0.8$.}
\end{figure}

The eigenvalues of the above reduced density matrix $\rho_A^{(k)}(k=1,2)$ can be calculated as (note that the eigenvalues of two reduced density matrices are equal),
\begin{equation}
\varepsilon_{1,2}^{(k)}=\frac{1}{2}(1\pm\eta),
\end{equation}
where the parameter $\eta$ is defined by
\begin{eqnarray}{\label{eq.20}}
\eta&=&\{c_3^2\cos^22\theta+\frac{1}{4}|f(t)\chi(t)|^4[2(c_1^2+c_2^2)\nonumber\\
& &+2(c_1^2-c_2^2)\sin(2\phi+\varphi)]\sin^22\theta\}^{1/2},
\end{eqnarray}
with
\begin{eqnarray}
\sin\varphi&=&\frac{f^2(t)\chi^2(t)+f^{\ast2}\chi^{\ast2}}{2|f(t)\chi(t)|^2},\nonumber\\
\cos\varphi&=&\frac{i(f^2(t)\chi^2(t)-f^{\ast2}\chi^{\ast2})}{2|f(t)\chi(t)|^2}.
\end{eqnarray}

Consequently, the von Neumann entropies $S(\rho_A^1(t))$ and $S(\rho_A^2(t))$ are given by
\begin{eqnarray}
&&S(\rho_A^1(t))=S(\rho_A^2(t))\nonumber\\
& &=-\frac{1-\eta}{2}\log_2{\frac{1-\eta}{2}}-\frac{1+\eta}{2}\log_2{\frac{1+\eta}{2}}\nonumber\\
& &=R(\eta),
\end{eqnarray}

Using Eq.~(\ref{eq.13}), the classical correlation can be written as
\begin{eqnarray}{\label{eq.23}}
\mathcal{C}(\rho_{AB}(t))&=&\max_{\{B_{k}\}}\{S(\rho_{A}) -
S(\rho_{AB}|{\{B_{k}\}})\nonumber\\
& &=1-\min_{\theta,\phi}\{\sum_kp_kS(\rho_k)\}\nonumber\\
& &=1-\min_{\theta,\phi}[R(\eta)].
\end{eqnarray}

Since the function $R(\eta)$ is a monotonically decreasing function, we can obtain the minimal value of $R(\eta)$ by choosing the suitable parameters $\theta$ and $\phi$ to ensure that the parameter $\eta$ defined in Eq.~(\ref{eq.18}) is maximal. From the Eq.~(\ref{eq.18}), it is not difficult to find that there exists an inequality
\begin{eqnarray}{\label{eq.24}}
\eta&\leq& \{c_3^2\cos^22\theta+\frac{|f(t)\chi(t)|^4}{4}[2(c_1^2+c_2^2)\nonumber\\
&&+2|c_1^2-c_2^2|]\sin^22\theta\}^{1/2}\nonumber\\
& \leq& \left\{ \begin{array}{cc}
|c_3| & \textrm{if $|c_3|>W(t)$}\\
W(t) & \textrm{if $|c_3|<W(t)$}
\end{array} \right.,
\end{eqnarray}
with
\begin{equation}
W(t)=\frac{|f(t)\chi(t)|^2}{2}\sqrt{2(c_1^2+c_2^2)+2|c_1^2-c_2^2|}.
\end{equation}
\begin{figure}
\centering
\includegraphics[width=7cm]{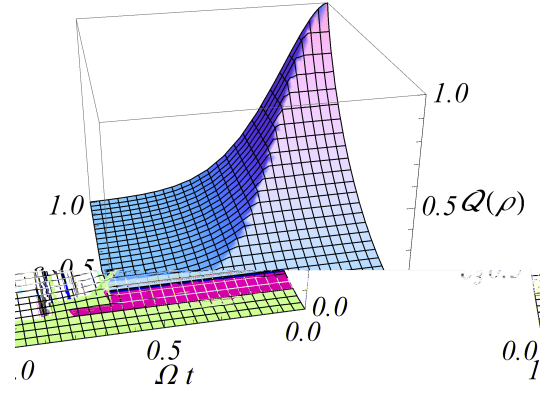}
\caption{\label{fig:fig3} The quantum discord of two atoms is plotted as a function of the dimensionless scaled time $\Omega t$ and parameter $c_3$ with $\alpha=1$ and $k/\Omega=0.05$.}
\end{figure}

Combining Eqs.~(\ref{eq.23}) and (\ref{eq.24}), we can rewrite the classical correlation as
\begin{equation}{\label{eq.27}}
\mathcal{C}(\rho_{AB}(t))=\sum_{j=1}^2\frac{1+(-1)^{j}m(t)}{2}\log_2[1+(-1)^jm(t)].
\end{equation}
where $m(t)=\max\{|c_3|,W(t)\}$.
\begin{figure}
\centering
\includegraphics[width=7cm]{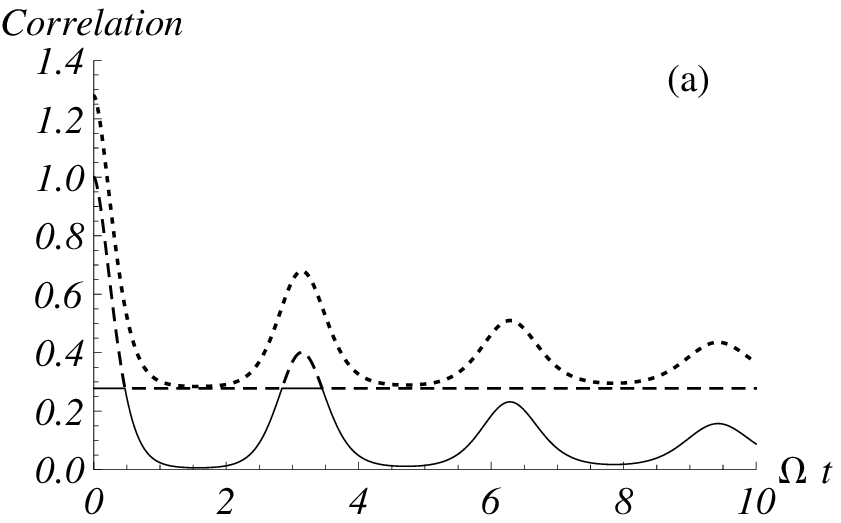}
\includegraphics[width=7cm]{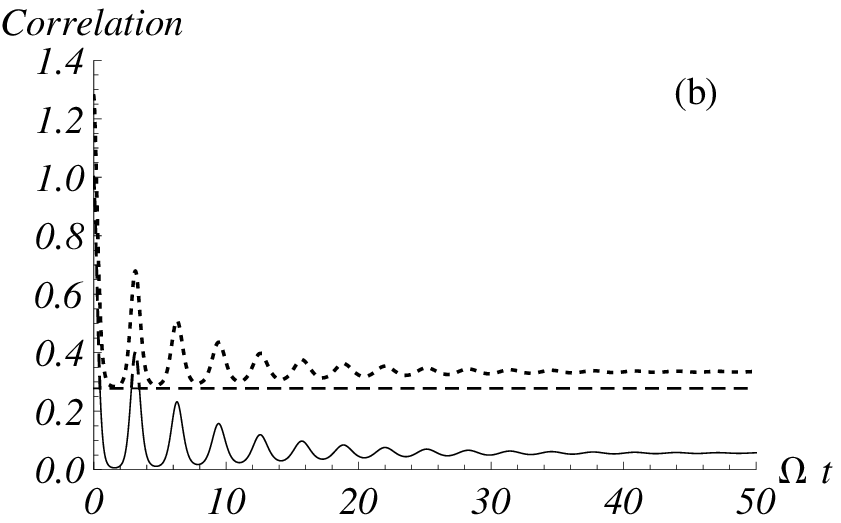}
\caption{\label{fig:fig4}Time evolution of quantum mutual information (dotted line), classical correlation (dashed line) and quantum discord (solid line) of two atoms as a function of the dimensionless scaled time $\Omega t$ with $k/\Omega=0.05$, $\alpha=0.8$ and $c_3=0.6$. Panel (a): $0\leq\Omega\leq10$. Panel (b): $0\leq\Omega\leq50$.}
\end{figure}

Then the quantum discord between two atoms can be expressed as
\begin{eqnarray}{\label{eq.28}}
&&\mathcal{Q} (\rho_{AB}(t))=\mathcal{I} (\rho_{AB}(t))-\mathcal{C} (\rho_{AB}(t))\nonumber\\
& =&2+\sum_{i=1}^4\lambda_i\log_2\lambda_i\nonumber\\
& &-\sum_{j=1}^2\frac{1+(-1)^{j}m(t)}{2}\log_2[1+(-1)^jm(t)].
\end{eqnarray}

We first focus on the X-type quantum states with state parameters $c_1=1$, $c_2=-c_3$ and $|c_3|<1$. Substituting these initial conditions into the Eqs.~(\ref{eq.16}), (\ref{eq.27}) and (\ref{eq.28}), the explicit expressions of the quantum mutual information, classical correlation and quantum discord can be obtained as,
\begin{eqnarray}
&&\mathcal{I} (\rho_{AB}(t))=\frac{1}{2}(1+c_3)\log_2(1+c_3)+\frac{1}{2}(1-c_3)\nonumber\\
&&\log_2(1-c_3)+\frac{1}{2}(1+|f(t)\chi(t)|^2)\log_2(1+|f(t)\chi(t)|^2)\nonumber\\
&&+\frac{1}{2}(1-|f(t)\chi(t)|^2)\log_2(1-|f(t)\chi(t)|^2),\nonumber\\
&&\mathcal{C}(\rho_{AB}(t))=\sum_{j=1}^{2}\frac{1+(-1)^jm(t)}{2}\log_2[1+(-1)^jm(t)],\nonumber\\
&&\mathcal{Q}(\rho_{AB}(t))=\mathcal{I}(\rho_{AB}(t))-\mathcal{C} (\rho_{AB}(t)).
\end{eqnarray}
where $m(t)=\max\{|c_3|,|f(t)\chi(t)|^2\}$.

In Fig.~\ref{fig:fig2}, we plot the quantum discord (solid line), the classical correlation (dashed line) and the quantum mutual information (dotted line) as a function of the dimensionless scaled time $\Omega t$ for two different values of $\alpha$ with $k/\Omega=0.05$ and $c_3=0.6$. It is shown that the phenomenon of the sudden transition between classical and quantum decoherence appears in this system within a short interaction time. Comparing the panel (a) with panel (b), we find that the quantum discord between two atoms is not destroyed by the dissipation of the cavities for a finite time interval and the transition time $\overline{\Omega t}$ can be tuned by adjusting the value of the parameter $\alpha$.
\begin{figure}
\centering
\includegraphics[width=7cm]{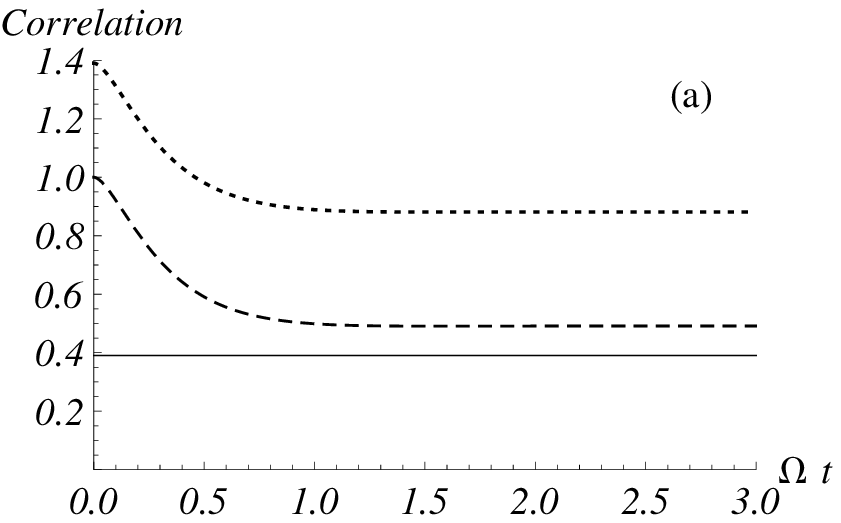}
\includegraphics[width=7cm]{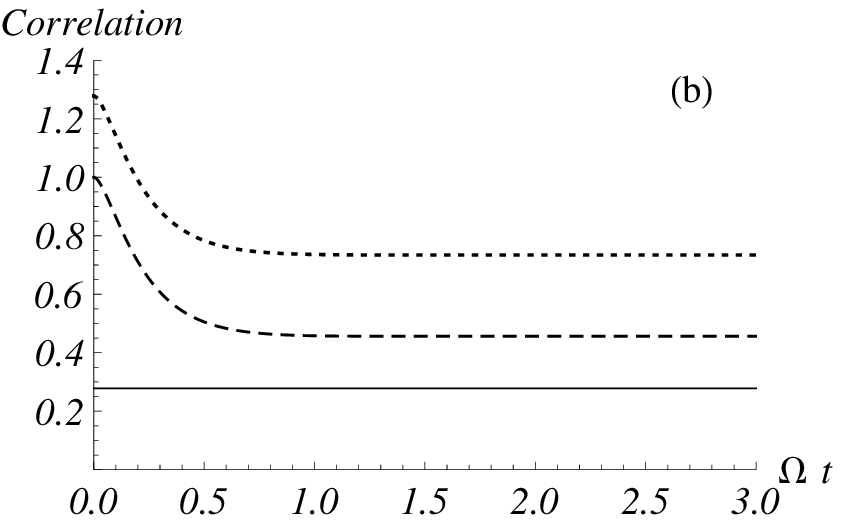}
\caption{\label{fig:fig5} Time evolution of quantum mutual information (dotted line), classical correlation (dashed line) and quantum discord (solid line) of two atoms as a function of the dimensionless scaled time $\Omega t$. Panel (a): $\alpha=0.8$, $c_3=0.7$ and $k/\Omega=2$. Panel (b): $\alpha=1.2$, $c_3=0.6$ and $k/\Omega=3$. }
\end{figure}

In order to illustrate the dependence of transition time on the initial states, the quantum discord $\mathcal{Q}(\rho)$ is displayed as a function of the dimensionless scaled time $\Omega t$ and the parameter $c_3$ with $\alpha=1$ and $k/\Omega=0.05$ in Fig.~\ref{fig:fig3}. We can see that the quantum discord is a constant over the time interval $\Omega t <\overline{\Omega t}$ and can be increased by decreasing the value of $c_3$. This consequence is in agreement with the conclusion which is obtained by Ref.~ \cite{17}.
\begin{figure}
\centering
\includegraphics[width=7cm]{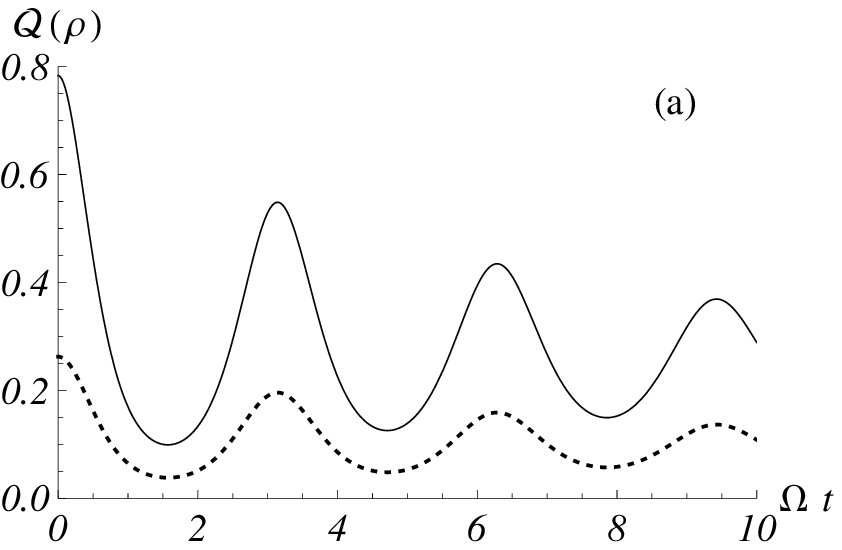}
\includegraphics[width=7cm]{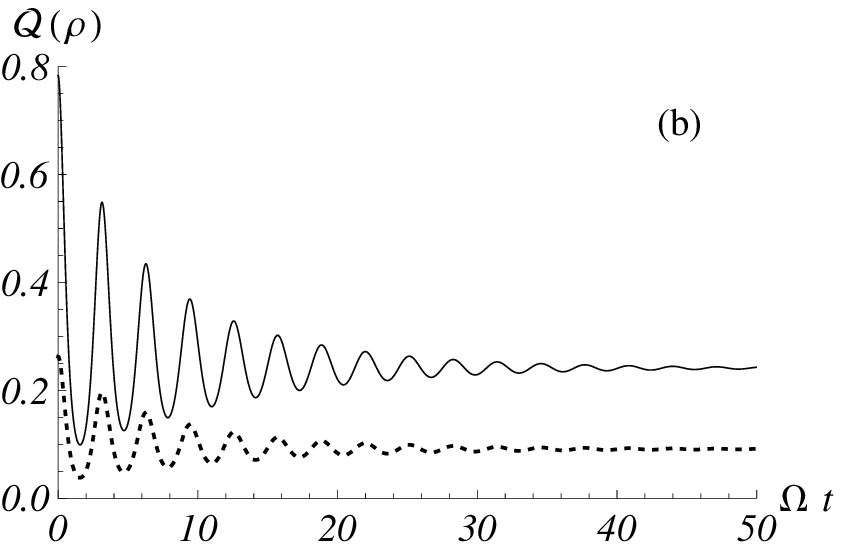}
\caption{\label{fig:fig6} Time evolution of quantum discord of two atoms as a function of the dimensionless scaled time $\Omega t$ with $\alpha=0.5$ and $k/\Omega=0.05$ for $r=0.5$ (dotted line) and $r=0.9$ (solid line). Panel (a): $0\leq\Omega t\leq10$. Panel (b): $0\leq\Omega t\leq50$. }
\end{figure}

Since the decay of quantum and classical correlations in this system corresponds to hyperexponential decay, the long-time behavior is different from the situation of exponential decay of Ref.~\cite{17}. In Fig.~\ref{fig:fig4}, we plot the correlations as a function of the dimensionless scaled time $\Omega t$ ($0\leq\Omega t \leq50$) with $\alpha=0.8$, $k/\Omega=0.05$ and $c_3=0.6$. It is quite clear that the quantum correlation revivals to a stable value after damping oscillation, which means that the initial quantum discord of two atoms can be partially preserved even when they are put into the two spatially separated dissipative cavities, respectively. The amount of stationary quantum discord depends on the initial state of the two atoms and the parameter $\alpha$.

The influence of the ratio $k/\Omega$ on the dynamics evolution of the quantum discord (solid line), classical correlation (dashed line) and quantum mutual information (dotted line) is displayed in Fig.~\ref{fig:fig5}. Comparing the Fig.~\ref{fig:fig5} with the Fig.~\ref{fig:fig4}, we can see that the quantum correlation of two atoms is completely unaffected by the decoherence of the cavities if we choose the suitable ratio $k/\Omega$. The reason is that the cavity fields deplete quicker with increasing the dissipation parameter $k$, the chance for atoms to interact with the cavity fields decreases. This result may have some applications in the quantum information processing and quantum memory.
\begin{figure}
\centering
\includegraphics[width=7cm]{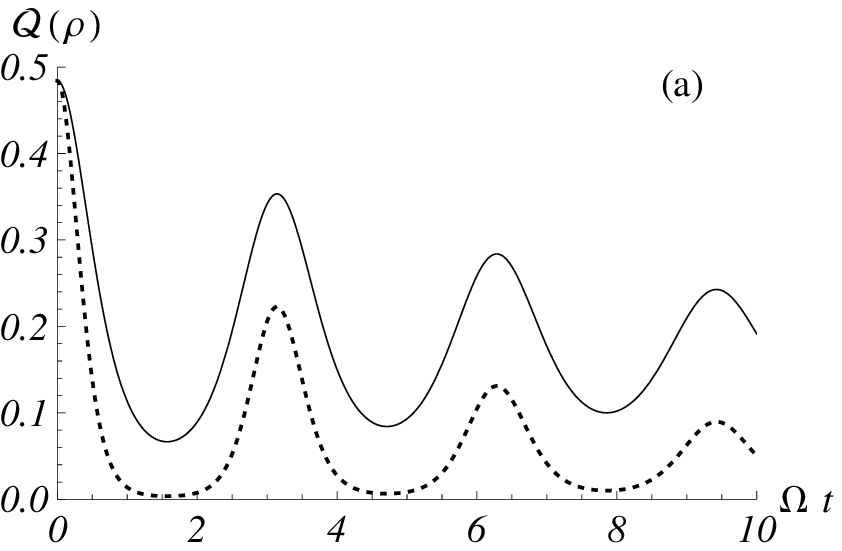}
\includegraphics[width=7cm]{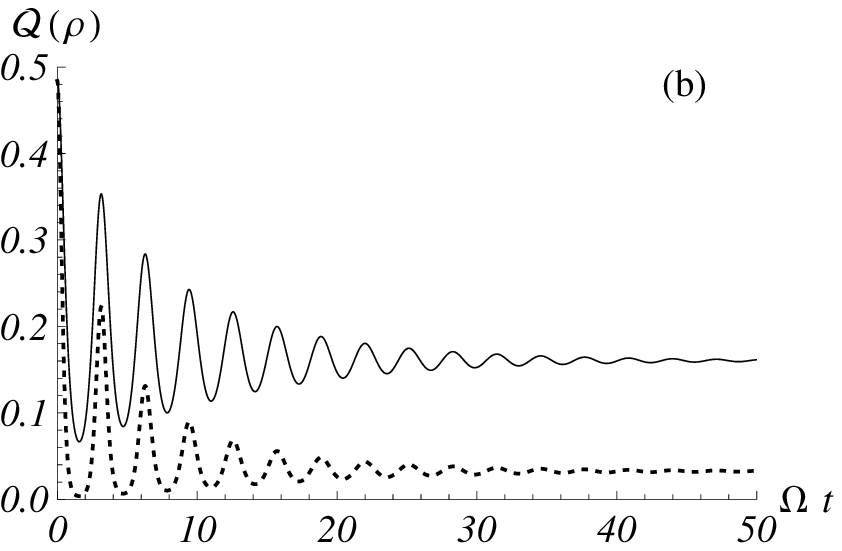}
\caption{\label{fig:fig7} Time evolution of quantum discord of two atoms as a function of the dimensionless scaled time $\Omega t$ with $r=0.7$ and $k/\Omega=0.05$ for $\alpha=0.8$ (dotted line) and $\alpha=0.5$ (solid line). Panel (a): $0\leq\Omega t\leq10$. Panel (b): $0\leq\Omega t\leq50$. }
\end{figure}

In the following, we consider the class of initial states for which $c_1=c_2=c_3=-r$, with $r\in[0,1]$. In this situation, the initial state $\rho_{\textrm{atom}}(0)$ turns out to be the Werner state \cite{23}
\begin{equation}
\rho_{\textrm{atom}}(0)=(1-r)\frac{I}{4}+r|\Psi^-\rangle\langle\Psi^-|,
\end{equation}
with
\begin{equation}
|\Psi^-\rangle=\frac{1}{\sqrt{2}}(|eg\rangle-\langle ge|).
\end{equation}

Using the similar procedure, we can obtain the quantum mutual information as
\begin{eqnarray}
\mathcal{I}(\rho_{AB}(t))&=&2+\sum_{i=1}^4\lambda_i\log_2\lambda_i,\nonumber\\
\lambda_1&=&\lambda_2=\frac{1-r}{r},\nonumber\\
\lambda_3&=&\lambda_4=\frac{1}{4}(1+r\pm2r|f(t)\chi(t)|^2),
\end{eqnarray}
the classical correlation is given by
\begin{eqnarray}
\mathcal{C}(\rho_{AB}(t))&=&\sum_{j=1}^2\frac{1+(-1)^jn(t)}{2}\log_2[1+(-1)^jn(t)],\nonumber\\
n(t)&=&\max\{r,r|f(t)\chi(t)|^2\},
\end{eqnarray}
and the quantum discord is
\begin{eqnarray}
&&\mathcal{Q}(\rho_{AB}(t))=\mathcal{I}(\rho_{AB}(t))-\mathcal{C}(\rho_{AB}(t))\nonumber\\
&&=\frac{1-r}{2}\log_2(1-r)+\sum_{j=1}^{2}\frac{1+r+(-1)^j2r|f(t)\chi(t)|^2}{2}\nonumber\\
&&\log_2[1+r+(-1)^j2r|f(t)\chi(t)|^2]-\sum_{j=1}^{2}\frac{1+(-1)^jn(t)}{2}\nonumber\\
&&\log_2[1+(-1)^jn(t)].
\end{eqnarray}

In Fig.~\ref{fig:fig6}, we plot the quantum discord $\mathcal{Q}(\rho)$ as a function of the dimensionless scaled time $\Omega t$ with $\alpha=0.5$ and $k/\Omega=0.05$ for $r=0.5$ (dotted line) and $r=0.9$ (solid line). It is shown that the phenomenon of sudden transition does not appear in this situation. Instead, the quantum discord revivals to a stable value after damping oscillation, which means that there is a stationary quantum discord between two atoms as the time approach to infinite. Furthermore, we also find that stationary quantum discord of two atoms increases with increaseing the parameter r.
\begin{figure}
\centering
\includegraphics[width=7cm]{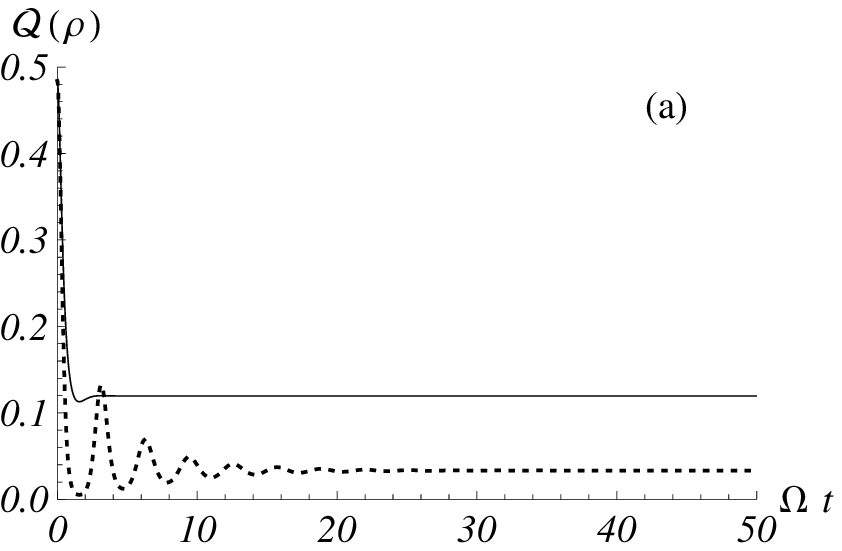}
\includegraphics[width=7cm]{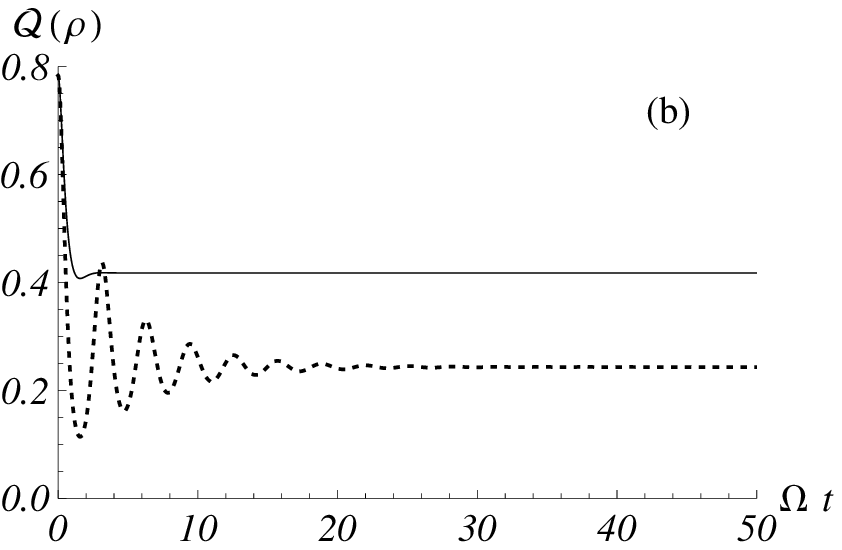}
\caption{\label{fig:fig8} Time evolution of quantum discord of two atoms as a function of the dimensionless scaled time $\Omega t$ for $k/\Omega=0.1$ (dotted line) and $k/\Omega=1$ (solid line). Panel (a): $r=0.7$ and $\alpha=0.8$. Panel (b): $r=0.9$ and $\alpha=0.5$. }
\end{figure}

The evolution of the quantum discord of two atoms is plotted with $r=0.7$ and $k/\Omega=0.05$ for $\alpha=0.8$ (dotted line) and $\alpha=0.5$ (solid line) in Fig.~\ref{fig:fig7}. It is obvious that there is a stationary quantum discord between two atoms and can be increased by decreasing the value of the parameter $\alpha$.

In order to show the influence of ratio $k/\Omega$ on the behavior of quantum discord of two atoms, we plot the quantum discord as a function of the dimensionless scaled time $\Omega t$ for two different values of the ratio $k/\Omega$ in Fig.\ref{fig:fig8}. We can see clearly from Fig.~\ref{fig:fig8} that for the different initial conditions, the quantum discord of the stationary state can be enhanced by increasing the value of the ratio $k/\Omega$.

\section{Conclusions}

In this paper, we investigate the phenomenon of sudden transition between classical and quantum decoherence in the study of quantum discord for the dissipative cavity QED system, which consists of two noninteracting two-level atoms, each trapped in a dissipative cavity. Firstly, we assume that the two atoms are prepared initially in the X-type quantum states with state parameters $c_1=1$, $c_2=-c_3$ and $|c_3|<1$. It is shown that the quantum discord of two atoms is not destroyed by the dissipation of the cavities for a finite time interval and partially recovers its initial values for a long interaction time, which means that there is a stationary quantum discord between the two atoms as the time approach to infinite. This demonstrates that the initial quantum discord of two atoms can be partially preserved even when they are put into the two spatially separated dissipative cavities,
respectively. Furthermore, we notice that the transition time $\overline{\Omega t}$ depends on the parameter of initial states of the two atoms and the mean photon number of the coherent field. Particularly, it is interesting to point out that the quantum discord of two atoms are completely unaffected by the dissipation of the cavities if we choose the suitable value of the ratio $k/\Omega$, which depends on the decay rate of two cavities and the atoms-field coupling constant. Then, we find that the sudden transition phenomenon does not appear in this system if the two atoms are initially in the Werner state. Instead, the stationary quantum discord of the two atoms still exists in the long-time regime. It is worth noting that the amount of the stationary quantum discord between the two atoms can be enhanced by increasing $k/\Omega$. These results may have potential applications in the quantum information processing and quantum memory.

\begin{acknowledgments}
This project was supported by the National Natural Science Foundation of China (Grant No. 10774131 and 11074310).
\end{acknowledgments}

\end{document}